\documentclass[aps,showpacs,showkeys,superscriptaddress]{revtex4}

\begin{document}
\title{Electronic properties of disclinated flexible membrane beyond the
inextensional limit: Application to graphene}

\author{E.A. Kochetov}\email{kochetov@theor.jinr.ru}
\affiliation{Bogoliubov Laboratory of Theoretical Physics, Joint
Institute for Nuclear Research, 141980 Dubna, Moscow region, Russia}

\author{V.A. Osipov}\email{osipov@theor.jinr.ru}
\affiliation{Bogoliubov Laboratory of Theoretical Physics, Joint
Institute for Nuclear Research, 141980 Dubna, Moscow region, Russia}

\author{R. Pincak}\email{pincak@saske.sk}
\affiliation{Bogoliubov Laboratory of Theoretical Physics, Joint
Institute for Nuclear Research, 141980 Dubna, Moscow region, Russia}
\affiliation{Institute of Experimental Physics, Slovak Academy of Sciences,
Watsonova 47,043 53 Kosice, Slovak Republic}


\begin{abstract}

Gauge-theory approach to describe Dirac fermions on a disclinated flexible
membrane beyond the inextensional limit is formulated. The
elastic membrane is considered as an embedding of $2D$ surface
into $R^3$. The disclination is incorporated through an $SO(2)$
gauge vortex located at the origin, which results in a metric with a
conical singularity. A smoothing of the conical singularity is
accounted for by replacing a disclinated rigid plane membrane with
a hyperboloid of near-zero
curvature pierced at the tip by the $SO(2)$ vortex. The embedding parameters
are chosen to match the solution to the von Karman equations. A
homogeneous part of that solution is shown to stabilize the theory.
The modification of the Landau states and density of
electronic states of the graphene membrane due to elasticity is discussed.
\end{abstract}

\maketitle

\section{Introduction}

Is is now generally accepted that the low-lying electronic states in
graphene can be accurately described by two-dimensional massless
Dirac fermions on a plane~\cite{review}. In experiment, multiform graphene
structures were observed thus stimulating studies of Dirac fermions
on curved graphene sheets (see, e.g.,~\cite{cortijo,pachos}). This
problem is markedly complicated when the curvature itself is
generated by topological defects like disclinations. Indeed, a
disclination is known in elasticity theory as a line defect which
can be produced by "cut and glue" Volterra process, namely, by
inserting or removing a wedge of material with the following gluing
of the dihedral sides. This immediately generates additional large
elastic strains inside the crystal. For flexible membranes, however,
there is a chance to screen out the strain field by buckling into a
cone. The problem thus reduces to coupling Dirac spinors to a
topologically non-trivial curved background.

The topological lattice defects in graphene are pentagons that
are equivalent to wedge disclinations. The first experimental
observation of a pentagon at the apex of a cone was provided by An
et al. in~\cite{an}. They used the scanning tunneling microscope
to study the structure of a conical protuberance and found five
bright spots at the apex of the nanocone. This was the first clear
evidence that the pentagon is located at the apex.
The bright spots indicate also that there is an enhanced charge density localized at
each carbon atom in the pentagon, which implies an increase in the
electronic density of states (DOS). This finding was confirmed by
the numerical tight-binding calculations for nanocones with
different number of pentagons at the apex~\cite{charlier}.
Explicit  manifestations of the topological effects in the electronic properties
of disclinated rigid graphene surfaces have been discussed at length in the literature
~\cite{jose92,jose93,lammert,jetplet00,jetplet01,pachos2,ando,jackiw,zaanen,sitenko}.

However, disclinations are the sources of long range elastic
stresses which also modify the electronic states. The
strain-induced effects in the electronic structure of graphene are
now a topic of intense research in strain engineering to produce
basic elements for all-graphene electronics. In particular,
designed strains can generate electron beam collimation, confined
states, quantum wires ~\cite{pereira} as well as energy gaps and a
zero-field quantum Hall effect~\cite{guinea} in graphene. To
incorporate internal stresses due to topological defects one is
supposed to go beyond the inextensional limit. Study of electronic
properties of disclinated flexible membranes beyond the
inextensional limit is thus of importance to address this new
field of research.

According to Volterra process disclination can be considered as a
conical singularity like strings in cosmology. The relevant
background is the curved spacetime where all the curvature is
concentrated at the apex of the cone. The metric in the
cylindrical coordinates is written as
\begin{equation}
ds^2 =dt^2+dz^2+dr^2+\alpha^2r^2d\varphi^2. \label{metric}
\end{equation}
Here the parameter $\alpha$ is related to the angular sector that
is removed or inserted to form the defect. In this case, any
attempt to build a closed loop around the disclination line will
result in a closure failure. The deficit angle is equal to
2$\pi\alpha$ with $\alpha=1-\nu$ where $\nu$ is the Frank index,
the basic topological characteristic of the disclination. The
positive sign of $\nu$ corresponds to the removing of a sector. In
this case the space has positive curvature. Correspondingly, for
negative $\nu$ one has a cone of negative curvature. Eventually, the
problem reduces to a Dirac equation in the curved
spacetime.
By the change $r\to r^{\alpha}/\alpha$ metric~(\ref{metric})
can be brought into the form
\begin{equation}
ds^2 =dt^2+dz^2+r^{2(\alpha-1)}(dr^2+r^2d\varphi^2) \label{metric2}
\end{equation}
which describes a featureless cosmic string located at the origin.

In spite of the elegant form of this approach, there is yet an
important open question concerning the so-called core region of the
defect. To the best of our knowledge, this problem was for the first time
raised in cosmological models~\cite{allen1,allen2} where
long-range effects of cosmic string cores were studied. In geometric
theory of defects, an influence of a disclination core on the
localization of electrons and holes was investigated
in~\cite{ribeiro}. In both cases, the tip of the conical singularity
is replaced by a smooth cap while at large distances a typical cone
with the deficit angle 2$\pi\alpha$ emerges. In cosmological models
the curvature of an infinite string is confined within a
cylinder of a small radius $a$ (the core radius) that possesses a
direct physical meaning: the string has a characteristic core
radius given by $a\approx 1/M$, where $M$ is a mass scale at which the string
is formed. Accordingly, the relevant $2D$ piece of the metric can be taken in the
form
\begin{equation}
ds^2 =  P^2(r/a)dr^2+\alpha^2r^2d\varphi^2, \label{coremetric}
\end{equation}
where the range of the angular coordinate is $\varphi\in [0, 2\pi)$ and
$P(r/a)$ is a smooth monotonic function satisfying the conditions
\begin{equation}
\lim_{r/a\rightarrow 0} P(r/a) = \nu, \qquad  P(r/a)=1, \quad r > a.
\label{P}
\end{equation}
For example, in~\cite{ribeiro}
the so-called flower-pot model was considered when the curvature of the disclinated
media is concentrated on a ring of radius $a$, which results in the formation of a "seam"
on the cylinder.

This approach is of interest in the description of linear defects
with a certain interior structure (finite thickness of a string).
However, the situation is more subtle for a disclination on an elastic $2D$ surface,
e.g., on a graphene sheet.
The specificity of  elastic membranes lies in that they may change both their
intrinsic and extrinsic geometries due to stretching and bending of the membranes, respectively.
This occurs because a flexible membrane can relieve the internal strain by buckling
out of the plane into a cone.
In the inextensional limit of the infinite rigidity, the stretching energy of a membrane of the radius
$R$ is proportional to the system size,
$R^2$, whereas a buckled membrane has the energy proportional to $\log R$.
In this case a creation of a disclination
by using the "cut and glue" process results in a true cone. This corresponds to a
point-like disclination defect. In reality, however, a membrane possesses finite elasticity,
so that one needs to go beyond the inextensional-limit approximation.

As is known, the classical theory of elasticity introduces a
characteristic velocity, but does not lead to a characteristic length.
In defect theory the length scale is introduced
phenomenologically through the core radius of the defect which
appears as an external parameter. Accordingly, there is no room within the linear theory
of elasticity for the description of the core region.
In the rotationally symmetric case it represents a small disc with a certain radius $r_0$.
Physically, the parameter $r_0$ comes into play through the boundary conditions imposed
on the stress tensor to insure that the internal strain inside a membrane be kept finite
to prevent it from falling apart.
It is this quantity $r_0$ that sets a relevant short-range length scale in the present problem
(similar to $a$ for cosmic strings).

In this paper, we attempt a variant of the
self-consistent gauge-theory approach to
account for both the smoothed apex and the
topological characteristic of the disclination defect.
To take into account finite elasticity,
we go beyond the inextensional--limit approximation.
In doing so, we invoke both the linear elasticity theory
as well as a phenomenological approach to account for the elastic
deformations.
Dynamical variables in our theory are the embeddings
of a $2D$ elastic surface into $R^3$.
Parameters of the embedding appear as the matter fields
interacting with an external gauge potential that describes a
disclination.
Within the linear scheme the model recovers the von Karman equations
for membranes with a disclination-induced source generated by
the gauge field.
We explicitly show that the elastic deformations emerge due to the defect.
We compute the effective metric generated by the disclination.
That metric is determined by the external gauge field and the dynamical embeddings.
It contributes to a topologically nontrivial part of the spin connection that
couples Dirac fermions to a $2D$ disclinated elastic manifold.

The paper is organized as follows.
Section II contains a brief
account of a gauge-theory approach to describe fermions on a disclinated surface
in the inextensional limit.
To take into account elastic deformations, a dynamical theory beyond the inextensional-limit approximation
is presented in Section III.
Within that theory  Dirac fermions couple to the underlying effective metric
generated by the disclination defect in the presence of elasticity.
In Section IV the derived
approach is specified to study electronic  properties of disclinated graphene.

\section{Infinite rigidity}

In the present Section we show that a disclination on a $2D$ manifold can be thought of as arising due
to an explicit breaking of the local rotational symmetry. This breaking can be enforced by a topologically
nontrivial external gauge field that carries the disclination "charge", $\nu$.
To set the stage, we begin with a
brief discussion of the case of an infinitely rigid membrane free to buckle into a cone in $R^3$.
This corresponds to the inextensional limit to be discussed from the elasticity theory point of view
in the next Section. Here we just show that the conical-type singularity generated by the vortex
results in a topological nontrivial contribution to the spin connection for Dirac fermions.
Our discussion essentially follows that of our earlier short
Communication~\cite{jetplet10}, except that it is extended now to include a disclination dipole
as well.

\subsection{Disclination vs gauge field}

Let $x^a$ be a set of local coordinates
on a Riemannian surface $\Sigma_0$.
(Indices $a,b,c,...=1,2$ are
tangent to $\Sigma_0$, whereas $i,j,k,...=1,2,3$
run over the basis of $R^3$). To describe this we find
it convenient to introduce an embedding
$\Sigma_0\to R^3$ that can be realized in terms of a $R^3$-valued
function $R_{(0)}^i(x^1,x^2)$. As the point
$(x^1,x^2)$ is varied, vector $\vec R_{(0)}$ sweeps the surface
$\Sigma_0$. This is nothing but a familiar two-parametric
representation of surfaces in $R^3$.
In what follows the function $R^i_{(0)}(x^1,x^2)$ is chosen to
specify an initial configuration $\Sigma_0$.
Representation for the induced metrics follows immediately
\begin{eqnarray}
g^{(0)}_{ab}&\equiv&(g_{\Sigma_0})_{ab}=
\partial_a\vec R_{(0)}\cdot
\partial_b\vec R_{(0)}.
\label{01.1}\end{eqnarray}

Equation (\ref{01.1}) is invariant under global $SO(3)$ rotations
of the vector $\vec R_{(0)}$.
To incorporate disclinations one should promote this invariance to a
local one. To this end,
consider the $R^3_{(0)}$-bundle over
$\Sigma_0$ with the structure group $SO(3)$. The $so(3)$ valued one
form $ (\vec W^{(0)}_a\cdot\vec L) dx^a$
serves as a connection one-form in the $R^3_{(0)}$-bundle
space over $\Sigma_0$, with $\vec W^{(0)}_a$ being the gauge potentials.
Here $L^i\in so(3)$ are the generators of the group.
By replacing in (\ref{01.1}) ordinary derivatives
$\partial_a\vec R_{(0)}$ by the covariant ones $\nabla_a\vec
R_{(0)}=\partial_a\vec R_{(0)} + [\vec W^{(0)}_a,\vec R_{(0)}]$,
one arrives at the locally $SO(3)$ invariant
representation for the induced metric,
\begin{eqnarray}
g_{ab}&=& g^{(0)}_{ab}(\vec W^{(0)})=\nabla_a\vec R_{(0)}\cdot\nabla_b\vec R_{(0)}=
\partial_{a}\vec R_{(0)}\cdot
\partial_{b}\vec R_{(0)}+
\partial_{a}\vec R_{(0)}[\vec W^{(0)}_{b},\vec R_{(0)}] +
\partial_{b}\vec R_{(0)}[\vec W^{(0)}_{a},\vec R_{(0)}] \nonumber \\
&+&(\vec W^{(0)}_{a} \vec W^{(0)}_{b})\vec R_{(0)}^2 - (\vec W^{(0)}_{a}\vec R_{(0)})(\vec
W^{(0)}_{b}\vec R_{(0)}).
\label{01.2}\end{eqnarray}
Topological disclinations can then be considered as arising
due to explicit breaking of the local rotational symmetry
by a fixed topologically nontrivial gauge potential
that generates a new metric as follows from Eq. (\ref{01.2}).
>From now on a metric induced due to either the gauge field or elastic deformations will be
denoted by $g_{ab}$ to reserve the symbol $ g^{(0)}_{ab}$ for the metric
tensor on the undisturbed surface, $\Sigma_0$.

In general case, a nonabelian gauge field $\vec W^{(0)}$ emerges to describe
disclinated surface . However, throughout this paper
we are primarily interested  in
the case $\Sigma_0=R^2$ so that only the $z$-component of the gauge field matters.
To illustrate this, consider a disclination defect placed at the origin of a plane
that can be bent but cannot be stretched.
We have $\Sigma_0=R^2,$
so that $(x^1=x,\,x^2=y)\in R^2$.
In this case $W_{\mu}^{(0)i=1,2}=0$ and
$W_{\mu}^{(0)i=3}=W^{(0)}_{\mu}$.
A singular vortex-like potential
\begin{equation}
W^{(0)}_x=\nu y/r^2,\,W^{(0)}_y=-\nu x/r^2,\,  r=\sqrt{x^2+y^2}\ne 0
\label{01.3}\end{equation}
is supposed to describe a topological disclination with a strength $\nu$ located at $\vec r=0$.
This potential locally is a pure gauge,
$$W^{(0)}_{a}(\vec r) =-\nu\partial_{a}\tan^{-1}\frac{y}{x}.$$
However, for any counter $C$ encircling the origin one has
\begin{equation}
\oint_{C} W^{(0)}_a dx^a=-2\pi\nu\ne 0.
\label{01.4}\end{equation} Since the counter integral in Eq.(\ref{01.4}) is a gauge invariant quantity,
the field $W^{(0)}_{\mu}$
cannot be gauged away to zero due to the topological obstruction. This is why that field is referred to as a
topologically non-trivial one. A physically observable quantity associated with that gauge field is
a nonzero flux, $\Phi=-2\pi\nu$,
through an area bounded by the counter $C$. It does not depend on small continuous deformations of that area.
This flux instead characterizes the gauge potential globally: it determines the first Chern characteristic class
the gauge potential $W^{(0)}$ belongs to. An electron encircling the origin naturally acquires a topological phase
associated with that nontrivial flux: the Aharonov-Bohm phase which distinguishes the gauge potential $W^{(0)}$
from a trivial one.

In the polar coordinates
$(r,\varphi)\in R^2$ a plane can be regarded as an embedding
$$(r,\varphi)\to
(r\cos\varphi, \ r\sin\varphi, 0), \quad 0<r<\infty,\,
0\le\varphi<2\pi.$$
The gauge potential takes the form
\begin{equation}
W^{(0)}_r=0,\,\, W^{(0)}_{\varphi}=-\nu.
\label{01.5}\end{equation}
The components of the induced metric (\ref{01.2}) can be easily read
off
\begin{equation} g_{rr}= 1, \qquad
g_{\varphi\varphi}=\alpha^2r^2, \qquad g_{r\varphi}=g_{\varphi r}=0,
\label{01.55}
\end{equation}
Evidently, this is a metric of a cone (cf. (\ref{metric})) which
at $\nu=0$ goes over to a flat one.

In general case the potential
\begin{equation}
W^{(0);(\nu_1,,..,\nu_N)}_{a}(\vec r)= -\sum_i^N\nu^{(i)}\frac{\epsilon_{ab}(\vec r-\vec r_i)^{b}}
{|\vec r-\vec r_i|^2},
\label{01.6}\end{equation}
where $a,b=x,y$, is supposed to describe $N$ disclination defects with the strengths
$\nu^{(i)}$ located at the points with the coordinates $\vec r_i,\, i=1,..,N$. To specify to an important
in applications case of the
disclination dipole located on the $x$-axis, we put $\vec r_1=(-L,0),\, \vec r_2=(L,0)$
and $\nu^{(1)}=-\nu^{(2)}=\nu$.
The above equation then yields
\begin{equation}
W^{(0)}_x=\nu\frac{4xyL}{r_{+}^2r_{-}^2},\quad W^{(0)}_y=\nu\frac{2L(y^2-x^2+L^2)}{r_{+}^2r_{-}^2},
\label{01.7}\end{equation}
where $r_{\pm}^2=(x\pm L)^2+y^2$.
Substituting this expression in Eq.(\ref{01.2}) one can in principle compute the induced metric.
To illustrate this, we work out
an explicit representation of the metric induced by a disclination dipole in the asymptotic region $L/r\ll 1$.
Far away from the dipole Eq.(\ref{01.7}) reduces in the polar coordinates to
\begin{equation}
W^{(0)}_{\varphi}=-2\nu\varepsilon\cos\varphi +{\cal O}(\varepsilon^2),
\quad W^{(0)}_r=\frac{2\nu\varepsilon}{r}\sin\varphi + {\cal O}(\varepsilon^2)/r,\quad \varepsilon:=L/r\ll 1.
\label{01.8}\end{equation}
Our representation of the induced metric (\ref{01.2}) then gives
\begin{equation}
g_{rr}= 1+{\cal O}(\varepsilon^2), \quad g_{\varphi\varphi}=r^2(1-4\nu\varepsilon\cos\varphi +{\cal O}(\varepsilon^2)),
\quad g_{r\varphi}=g_{\varphi r}= 2\nu\varepsilon r\sin\varphi +r{\cal O}(\varepsilon^2),
\label{eq:01.9}\end{equation}
This metric describes the asymptotic representation of the line element
for a screw dislocation (see, e.g., \cite{putingam}),
with the Burgers vector $b^y=-4\pi\nu L$ being perpendicular to the defect region.
The gauge-theory approach is thus seen to recover a well-known result: at large distances
from the defect a disclination dipole can be thought of as a screw dislocation.

\subsection{Incorporating fermions}

Let us now move on to a problem of coupling fermions to
a given disclination.
As is known, the topologically
nontrivial gauge field reasserts itself in the Dirac equation as a
topologically nontrivial piece of the spin
connection~\cite{jetp}. That part of the connection carries a
topologically nontrivial flux that does not depend on smooth
continuous changes of the underlying metric due to small elastic
deformations.
To incorporate fermions on the $2D$ background $(\Sigma_0=R^2,\vec r\ne 0;\, g^{(0)}_{ab}(W^{(0)}))$
we need a set of orthonormal frames $\{e_{\alpha}(W^{(0)})\}$ which yield the same metric,
$g^{(0)}_{ab}(W^{(0)})$, related to each other by the local $SO(2)$ rotation,
$$e_{\alpha}\to e'_{\alpha}={\Lambda}_{\alpha}^{\beta}e_{\beta},\quad
{\Lambda}_{\alpha}^{\beta}\in SO(2).$$

It then follows that
$g_{ab} = e^{\alpha}_{a}e^{\beta}_{b} \delta_{\alpha
\beta}$ where $e_{\alpha}^{a}$ is the zweibein, with the
orthonormal frame indices being $\alpha,\beta=\{1,2\}$, and
coordinate indices $a,b=\{1,2\}$ (from now on we drop an explicit
$W$-dependence of the metric). As usual, to ensure that
physically observed values be independent of a particular choice of the
zweinbein fields, a local $so(2)$--valued gauge field
$\omega_{\mu}$ is to be introduced. The gauge field of the local
$SO(2)$ group is referred to as a spin connection. For the theory to be
self-consistent, zweinbein fields must be chosen to be covariantly
constant \cite{witten}:
$$\partial_{a}e^{\alpha}_{b}
-\Gamma^{c}_{ab}e^{\alpha}_{c}+(\omega_{a})^{\alpha}_{\beta}
e^{\beta}_{b}=0,$$ which determines the spin connection
coefficients explicitly
\begin{equation}
(\omega_{a})^{\alpha\beta}= e_{b}^{\alpha}D_{a}e^{\beta b},\quad D_{a}=\partial_{a}+\Gamma_{a},
\label{2.1}\end{equation}
with $\Gamma_{a}$ being the Levi-Civita connection.
The Dirac equation on a surface $(\Sigma_0,\, g^{(0)}_{ab}(W))$ is written as
\begin{equation}
i\gamma^{\alpha}e_{\alpha}^{a}(\partial_{a}+\Omega_{a})\psi=E\psi,
 \label{2.2}\end{equation}
with
\begin{equation}
\Omega_{a}=\frac{1}{8}\omega^{\alpha\ \beta}_{\ a}
[\gamma_{\alpha},\gamma_{\beta}] \label{2.3}\end{equation} being
the spin connection in the spinor representation. In two space dimensions,
the Dirac matrices can be chosen to be the Pauli matrices,
$\gamma_1=-\sigma_2, \quad \gamma_2=-\sigma_1$.
In the case under consideration
Eq.~(\ref{2.1}) gives
\begin{equation}
\omega^{12}_r=\omega^{21}_r= 0,\quad \omega^{12}_{\varphi}=
-\omega^{21}_{\varphi}=
1-\alpha.
\label{2.3}\end{equation}
Hence, topologically nontrivial gauge field (\ref{01.5})
results in a conical singularity of the spin connection.
The flux
$$\oint_{C} \omega^{12}_{\varphi}d\varphi=2\pi\nu\ne 0$$
represents a "net" effect produced by a disclination on the moving electrons.
We thus show that the gauge-field approach in the inextensional limit
exactly coincides with the standard "cut-and-glue" procedure.

However, a cone with a point-like apex is mathematical abstraction since in a real situation the media has a
finite stiffness, which would inevitably result in a certain smearing of a conical singularity.
Therefore, a proper description of the disclination implies a smooth deformation of the metric
and at the same time  one has to preserve a conical behavior far away from the origin.
Although such a surface can effectively be approximated by a hyperboloid, we show
now that one cannot incorporate finite elasticity into the theory by simply replacing a cone by a smooth surface that
asymptotically approaches a cone far away from the origin. This would simply eliminate the defect.

To illustrate this, consider an upper half of a hyperboloid as an
embedding
\begin{equation}
(\xi,\varphi)\to (a\,{\sinh\xi}\cos\varphi,a\,{\sinh\xi}\sin\varphi,
c\,\cosh\xi), \quad 0\le\xi<\infty, 0\le\varphi<2\pi.
\label{hyperb}\end{equation}
The components of the induced metric can be written as
\begin{equation}
g_{\xi\xi}=a^2\cosh^2\xi+c^2\sinh^2\xi,\quad g_{\varphi\varphi}=
a^2\sinh^2\xi,\quad g_{\varphi\xi}=g_{\xi\varphi}=0,
\label{2.4}\end{equation} which in view of (\ref{2.1}) gives for the
spin connection coefficients
\begin{equation}
\omega^{12}_{\xi}=\omega^{21}_{\xi}= 0,\quad \omega^{12}_{\varphi}=
-\omega^{21}_{\varphi}=
\left[1-\frac{a\,\cosh\xi}{\sqrt{g_{\xi\xi}}}\right]=:\omega(\xi).
\label{2.5}\end{equation}
The spin connection  in the spinor $SO(2)$ representation becomes
\begin{equation}
\Omega_{\varphi}=i\omega\sigma_3. \label{2.6}\end{equation} Since
$\omega(\xi)$ goes to zero as $\xi\to 0$ a circulation of that
field over a loop encircling the origin gives a flux which tends
to zero as the counter shrinks to zero,
$$\lim_{\epsilon\to 0}\oint_{C_{\epsilon}}\omega^{12}_{\varphi}d\varphi=0,$$
where $C_{\epsilon}$ stands for a closed counter which encloses a small area $\sim\epsilon^2$ around the origin.
This equation implies that there is no a topologically nontrivial part in the flux.
It is therefore clear that one should work out some other way to explicitly accommodate elastic deformations in
the "cut-and-glue" procedure that would preserve a conical singularity at the origin.

\section{Finite  rigidity}

In the present Section we formulate the linear elasticity theory
in terms of the embeddings, which proves convenient to incorporate
defects. We thus arrive at the von Karman equations to describe
elastic disclinated media away from the limit of infinite
rigidity. Our consideration allows then to introduce an important
notion of the defect core radius, $r_0$. We assume that the linear
theory works well in the region $r\ge r_0$ whereas it breaks down
within the core, at $r<r_0$. We then briefly discuss the
phenomenological approach discussed earlier to effectively account
for the elastic properties of disclinated
membranes~\cite{jetplet10}. Matching the linear von Karman theory
at the boundary point $r=r_0$ with the phenomenological approach
enables us to represent the phenomenological parameters in terms
of the elasticity and bending constants.

\subsection{Elastic surface}

Let us start by discussing the
elastic properties of a $2D$ manifold in the absence of defects.
Under elastic deformations a surface $\Sigma_0$ evolves into
some other Riemannian surface $\Sigma,$
which can be thought of as a diffeomorphic map, $\phi:\,\Sigma_0\to \Sigma$.
Again, we find it convenient
to introduce the embedding
$\Sigma\to R^3$ that can be realized in terms of a $R^3$-valued
function $R^i(x^1,x^2)$,
the point being that~\cite{jpa99}
\begin{equation}
\vec R(x):=\phi^*\vec R_{(0)}=\vec R_{(0)}[\phi(x)],
\label{3.1}\end{equation}
where $\phi^*$ is a pullback of $\phi:\,\Sigma_0\to\Sigma$.
The induced metrics becomes
\begin{eqnarray}
g_{ab}&\equiv&(\phi^*g_{\Sigma})_{ab}
=(g_{\Sigma})_{cd}\frac{\partial{\phi^c}}{\partial x^a}\cdot
\frac{\partial{\phi^d}}{\partial x^b}=\frac{\partial{\vec R}}
{\partial{\phi^c}}\cdot\frac{\partial{\vec R}}{\partial{\phi^d}}\,
\frac{\partial{\phi^c}}{\partial x^a}\cdot
\frac{\partial{\phi^d}}{\partial x^b}=
\partial_a{\vec R}\cdot\partial_b{\vec R},
\label{3.3}\end{eqnarray} where the set $\{\phi^a\}$ stands for
local coordinates on $\Sigma$. The strain tensor is then determined
to be
$$E_{ab}=g_{ab}-g^{(0)}_{ab}.$$

The properties of a fluctuating elastic surface are encoded in the action
\begin{equation}
F=F_{el}+F_{fl}, \label{3.2}\end{equation} where $F_{el}$ describes
the elastic properties of the media, whereas $F_{fl}$ stands for the
Helfrich-Canham action to describe the energy of a free fluctuating
surface. Explicitly the stretching energy is taken to be quadratic
in the strain,
\begin{equation}
F_{el}= -\frac{1}{8}\int_{\Sigma_0} dx^1dx^2\sqrt {g^{(0)}}\left\{\lambda(tr E)^2
+2\mu\, trE^2\right\},
\label{3.4}\end{equation}
where $tr E=g_{(0)}^{ab}E_{ab},\, g^{(0)}=det ||g^{(0)}_{ab}||$ and summation over repeated
indices is assumed. Here $\lambda$ and $\mu$ are the $2D$ Lame coefficients.
The stress tensor is then introduced to be
\begin{equation}
\sigma_{ab}=2\mu E_{ab}+\lambda tr E\delta_{ab}.
\label{stress}
\end{equation}

The Helfrich-Canham bending energy of a membrane depends on its mean curvature $H$ and
Gaussian curvature $K$ ~\cite{helf,canham},
\begin{equation}
F_{fl}=\frac{\kappa}{2}\int_{\Sigma_0}\,\sqrt{g^{(0)}} dx^1dx^2 \,H^2
+ \frac{\kappa_G}{2}\int_{\Sigma_0}\sqrt{g^{(0)}}\, dx^1dx^2 \,K
\label{3.5}\end{equation}
where $\kappa$ is a bare bending rigidity and $\kappa_G$ is a Gaussian
rigidity. $H
=g_{(0)}^{ab}K_{ab}$ is the mean (extrinsic) curvature, and
$K=det\,g_{(0)}^{ab}K_{bc}$ is referred to as the Gaussian (intrinsic)
curvature. Here
\begin{equation}
K_{ab}=\vec N\cdot D_{a}D_{b}\vec R \label{3.6}\end{equation} is
the curvature tensor, and $\vec N$ is the unit normal to the
surface $$\vec N=\frac{[\partial_1\vec R,\partial_2\vec R]}
{|[\partial_1\vec R,\partial_2\vec R]|}.$$ The covariant
derivative $$D_{a}:=\partial_{a}+\Gamma_{a}$$ includes the Levi-Civita
connection $\Gamma_{a}$.

\subsection{Disclinations in flexible membranes}

To incorporate disclinations originally distributed on $\Sigma_0$
one needs to make in Eq.(\ref{3.3}) substitution
$$\partial_a\vec R_{(0)}\to
\nabla_{a}\vec R_{(0)}=
\partial_{a}\vec R_{(0)} + [\vec W^{(0)}_{a},\vec R_{(0)}].$$
As a result, metric $g^{(0)}_{ab}$ goes over to $g_{ab}$
given by Eq.(\ref{01.2}).
To derive equations of motion that follow from the Hamilton's principle
of least action, $\delta F=0$,
we first need to specify the embedding $R^i(x^1,x^2)$. In plane elasticity theory, an elastic deformation
is represented by a displacement vector $u_x(x,y),u_y(x,y)$. If a membrane is allowed to buckle out of the plane,
we must add an extra function $f(x,y)$ to describe the "deflection".
We thus choose  the embedding in the following way
\begin{equation}
\vec R(x^1,x^2)=\vec R_{(0)}+\vec U,
\label{3.7}\end{equation}
where
$\vec R_{(0)}=(x,y,0)$ and $\vec U=(u_x,u_y,f(x,y))$ is a displacement of the $(x,y,0)$ point
under deformation. In the linear approximation we may omit the terms quadratic in
displacements $u_a$ as well as in the source strength $\nu$.
The strain tensor then becomes
\begin{eqnarray}
E_{ab}
&=& \partial_au_b+\partial_bu_a +\partial_af\partial_bf-
\epsilon_{\alpha a}W^{(0)}_bR_{(0)}^{\alpha} -
\epsilon_{\alpha b}W^{(0)}_aR_{(0)}^{\alpha} +{\cal O}(u^2,u\partial f, W^2).
\label{3.8}\end{eqnarray}
Introducing the Airy stress function $\chi$,
$$\sigma_{ab}=\epsilon_{ac}\epsilon_{bd}\partial_c\partial_d\chi,$$
one eventually gets the following equations of motion
\begin{eqnarray}
\kappa\Delta^2f = (\partial_y^2\chi)(\partial_x^2 f)+(\partial_x^2\chi)
(\partial_y^2 f)-
2(\partial_x\partial_y\chi)(\partial_x\partial_y f), \nonumber \\[0.2cm]
K_0^{-1}\Delta^2\chi = (\partial_x\partial_y f)^2-(\partial_x^2 f)
(\partial_y^2 f) - \epsilon_{ab}\partial_a W^{(0)}_b,
\label{3.9}\end{eqnarray}
where $K_0=4\mu(\lambda +\mu)/(\lambda +2\mu)$ is the $2D$ Young's modulus.

A single disclination located
at the origin of a plane is described by the potential (\ref{01.3}).
This results in
$$- \epsilon_{ab}\partial_a W^{(0)}_b=\nu\,\Delta \log r=2\pi\nu\delta (\vec r),$$
so that Eqs.(\ref{3.9})
are exactly the von Karman equations  for a defect in a flexible membrane~\cite{nelson}.
Note, however, that the source term
does not appear in (\ref{3.9}) {\it ad hoc} but is rather
generated by the gauge field due to a disclination.
In case $N$ disclinations with strengthes $\nu_i,\, i=1,2,...,N$ are located
at points $\vec r_i$ one should use the gauge
potential (\ref{01.6}). The source term in Eq.(\ref{3.9}) is then computed to be
$$ - \epsilon_{ab}\partial_a W^{(\nu_1,,..,\nu_N)}_{b}=
2\pi\sum^{N} _i\nu_i\delta (\vec r-\vec r_i).$$

Dynamically induced metric on $\Sigma$ takes the form
\begin{equation}
g_{ab}
= \delta_{ab}+\partial_au_b+\partial_bu_a +\partial_af\partial_bf.
\label{3.10}\end{equation}
Both functions $\chi$ and $f$ or, equivalently, $u_a$ and $f$ contribute to it.
Besides, function $f(x,y)$ determines a shape of the emergent surface.
To see this, let us reexamine the case of a disclination on a plane that can be
bent but cannot be stretched considered in Section II
(see Eq.(\ref{01.55})).
Within our approach it corresponds to the case of $K_0\to\infty$.
Since there is no in-plane stretching
one may put $u_{a}=0$ in Eqs.(\ref{3.9}).
The second von Karman equation takes the form
\begin{equation}
(\partial_x\partial_y f)^2-(\partial_x^2 f)
(\partial_y^2 f) =  -2\pi\nu\delta (\vec r),
\label{3.11}\end{equation}
whereas the induced metric becomes,
\begin{equation}
g_{ab}
= \delta_{ab}+\partial_af\partial_bf.
\label{3.12}\end{equation}
Equation (\ref{3.11}) possesses an obvious solution $f=\pm\sqrt{2\nu}r$,
which is a defining equation of a cone.
The metric (\ref{3.12}) coincides with that given by Eq.(\ref{01.55}) up to
${\cal O}(\nu^2)$ order as it should in the linear approximation.

\subsection{Solution to the von Karman equations}

It is instructive to reveal a geometrical structure of Eq.(\ref{3.11}).
The embedding (\ref{3.7}) tells us that for an infinite stiffness
the structure of the surface $\Sigma$ is entirely determined by function
$z=f(x,y)$. In particular, the Gaussian curvature of the surface equals
\begin{equation}
K=\frac{(\partial_x^2 f)(\partial_y^2 f)-(\partial_x\partial_y f)^2}{(1+(\partial_xf)^2+(\partial_yf)^2)^2}.
\label{3.13}\end{equation}
It is clear that a general solution to Eq.(\ref{3.11}) must scale as $f\sim \sqrt{\nu}$. In the linear approximation
one can therefore drop the $f$-dependent terms in the denominator in (\ref{3.13}), whereupon that equation
takes the form
\begin{equation}
K=2\pi\nu\delta (\vec r).
\label{3.14}\end{equation}
This means that all the curvature is located in this case at the apex.
Thus, to get the curvature spread over some finite area one needs to take
into consideration elastic properties of a media. In other words, one needs
to consider the whole
set of the von Karman equations (\ref{3.9}) at finite elasticity and bending constants.

To this end, let us first rewrite the von Karman equations in the dimensionless form.
Under the substitutions $\chi\to \chi\kappa, \, \vec r\to \vec rr_0$ and $f\to fr_0$
where $r_0$ is yet unspecified parameter with the dimension of length, those equations become
\begin{eqnarray}
\Delta^2f &=& (\partial_y^2\chi)(\partial_x^2 f)+(\partial_x^2\chi)
(\partial_y^2 f)-
2(\partial_x\partial_y\chi)(\partial_x\partial_y f), \nonumber \\[0.2cm]
\epsilon\Delta^2\chi &=& 2\pi\nu\delta (\vec r)-K,
\label{3.16}
\end{eqnarray}
where the parameter $\epsilon=\kappa/(K_0r_0^2)$ and
the Gaussian curvature $K=det (\partial_a\partial_b f).$
All the functions as well as the coordinates entering these equations are
now dimensionless. Since in $2D$ one gets $[K^{1/3}_0]=[\kappa]=E$, the parameter $\epsilon$ is dimensionless as well.
The inextensional limit amounts to that of $\epsilon \to 0$.
Let us denote the solutions to the von Karman equations in this limit by $f_0$ and $\chi_0$.
The second line in Eqs.(\ref{3.16}) becomes
$$(\partial_x^2 f_0)(\partial_y^2 f_0)-(\partial_x\partial_y f_0)^2=2\pi\nu\delta (\vec r)$$
which has an obvious solution $f_0=\pm\sqrt{2\nu}r$ that describes as already mentioned a true cone.
Inserting $f_0$ into the first equation of (\ref{3.16}) gives
\begin{equation}
1=(x^2\partial^2_{xx}+y^2\partial^2_{yy}+2xy \partial^2_{xy})\chi_0.
\label{3.17}\end{equation}
This equation possesses an obvious inhomogeneous solution $\chi_0=-\log r$
discussed at length in~\cite{nelson}. However, there is also a nontrivial solution
of the corresponding homogeneous equation
missed in ~\cite{nelson}. A general solution to Eq.(\ref{3.17}) must include a homogeneous term and
reads
\begin{equation}
\chi_0=-\log r+q r.
\label{3.18}\end{equation}
Here $q$ is an arbitrary constant. This becomes evident upon rewriting Eq.(\ref{3.17}) in
the polar coordinates, $1=r^2\partial^2_{rr}\chi_0.$ The homogeneous part of the solution (\ref{3.18})
turns out to be
of the utmost importance in stabilizing the theory and bringing out the physical meaning of the parameter $r_0$.

To see this, let us for a moment restore an explicit $r_0$ dependence in Eq. (\ref{3.18}):
\begin{equation}
\chi_0=-\log\frac{r}{r_0}+q\frac{r}{r_0}.
\label{3.19}\end{equation}
Consider an elastic thin disk of the radius $R$ with a disclination sitting at the origin.
Because of rotational symmetry,
$$\sigma_{r\varphi}=-\partial_r(\frac{1}{r}\partial_{\varphi}\chi)$$
vanishes identically for (\ref{3.19}).
The radial component of the strain tensor
$$\sigma_{rr}=\frac{1}{r}\partial_r\chi +\frac{1}{r^2}\partial^2_{\varphi\varphi}\chi$$
yields, however
\begin{equation}
\sigma_{rr}=-\frac{1}{r^2}+\frac{q}{rr_0}.
\label{3.20}\end{equation}
It is clear that $\sigma_{rr}(r=R)$ vanishes as $R\to \infty$. It is singular, however, at the origin.
Moreover, if we ignored the homogeneous solution in (\ref{3.20})
we would run into a serious problem. Namely, a typical way to remove the singularity
is to delete a small disk of material around the origin. It is easy to see that for $q=0$
the strain tensor $\sigma_{rr}$ behaves like $-1/r^2$ at the boundary of the excised disk. This is clearly
physically unacceptable, which signals the instability
of the theory (see discussion in~\cite{nelson}).
It should be stressed that the general solution (\ref{3.19}) allows us to avoid this difficulty.
To show this, let us delete a small disk of the radius $r_0$ around the origin and
require that $\sigma_{rr}(r=r_0) =\sigma_0$. This yields $q =1+\sigma_0r_0^2.$
If one requires vanishing stresses at the inner boundary, one should put $\sigma_0=0$.
Therefore, the parameter $r_0$ characterizes the core region of the disclination.

It is also important to note that the homogeneous term in the solution (\ref{3.20})
significantly affects the stretching energy of the membrane,
\begin{equation}
E_s=\frac{1}{2K_0}\int_{r> r_0}d^2\vec r\,(\nabla^2\chi)^2.
\label{3.21}\end{equation}
Since $\nabla^2\log r=2\pi\delta(\vec r)$ the stretching energy (\ref{3.21}) for the stress function (\ref{3.20})
with $q=0$ becomes an identical zero. This is exactly the conclusion reached in~\cite{nelson}. However,
this is not physically appropriate,
since this result should follow only
in the limit $K_0\to \infty$. If we instead compute (\ref{3.21}) at $q\neq 0$, we will get
\begin{equation}
E_s\propto\frac{\kappa^2q^2}{K_0r_0^2}\log \frac{R}{r_0}=\epsilon\kappa q^2\log \frac{R}{r_0},
\label{3.22}\end{equation}
which indeed vanishes as $K_0\to \infty$. It should be stressed that the same logarithmic behavior has the bending energy,
which is written as~\cite{nelson}
\begin{equation}
E_b = 2\pi\nu\kappa\log\frac{R}{r_0}.
\label{3.222}\end{equation}
At $q\sim 1$ one has $E_s\sim \epsilon E_b$ as it should be at small $\epsilon$.
Because of the fact that the entropy in the Kosterlitz-Thouless argument also increases logarithmically,
this result provides an interesting possibility of disclination-mediated
phase transitions that might be realized in $2D$ elastic membranes (see, e.g.,~\cite{nelson2}).

\subsection{Phenomenological approach vs von Karman equations}

Let us now turn back to the dimensionless set of the von Karman equations
(\ref{3.16}).  We seek a general solution in the form
\begin{equation}
f=\sum_{n=0}^{\infty}\epsilon^n f_n,\quad \chi=\sum_{n=0}^{\infty}\epsilon^n \chi_n,
\label{3.23}\end{equation}
where we have already found $f_0$ and $\chi_0$. These series are supposed to converge, provided
the linear approximation is valid. Inserting (\ref{3.23}) back into (\ref{3.16}) results in a set of the
self-consistent coupled equations to determine step by step the functions $f_n$ and $\chi_n$.
Technically, those equations for $n\ge 1$
turn out to be quite complicated. Their analysis will be given elsewhere.
To get some insight, we invoke instead a sort of phenomenological approach
to effectively incorporate elasticity in the Dirac equation discussed earlier in~\cite{jetplet10}. We match the
information that follows from our approach with that provided by the the von Karman equations.
This enables us to explicitly determine the dependence of the phenomenological
parameters on the elasticity and bending constants to
analyze the electronic properties of elastic graphene.

To begin with, let us briefly recall the phenomenological
theory~\cite{jetplet10}. As was already mentioned, a rigid plane
pierced by a vortex results in a conical singularity. Let us now
assume that the membrane possesses a small finite elasticity. In
that case the vortex will produce the singularity at the origin as before
and, additionally, it causes the medium to
respond by smoothing the conical shape due to elasticity. We suggest
that both effects can be taken into account by placing the
vortex on the tip of a hyperboloid of a near-zero curvature.
That sort of hyperboloid is supposed to
effectively emerge as a response of the elastic plane to a
disturbance caused by the defect at large distances, $r>r_0$. The
parameters of the hyperboloid must fulfil some natural requirements
to be formulated shortly. In this way we arrive at the effective
metric that takes into account a response of the elastic media to
the disturbance caused by the defect.

Explicitly, we employ the embedding (\ref{hyperb}), with dimensionless (scaled by $r_0$) parameters $a$ and $c$
that appear as the phenomenological parameters of the theory.
The gauge field (\ref{01.5}) represents the vortex at the origin. It induces the following
metric:
\begin{equation}
g_{\xi\xi}=a^2\cosh^2\xi+c^2\sinh^2\xi,
\quad g_{\varphi\varphi}= a^2\alpha^2\sinh^2\xi,\quad g_{\varphi\xi}=g_{\xi\varphi}=0,
\label{2.7}
\end{equation}
where $\alpha=1-\nu$ and $\nu$ is the declination charge. At $\nu=0$ this metric reduces to
that of a true hyperboloid given by~(\ref{2.4}).
Comparing the embedding (\ref{hyperb}) with Eq.(\ref{3.7}) tells us that $c$ must scale as $\sqrt{\nu}$, which seems
natural: buckling is induced by the source.
Let us further suggest that the elasticity coefficients are accumulated in
parameter $a$, the inextensional limit corresponding to $a\to \infty$. In other words, we should get that
$a\to \infty$ as $K_0\to \infty$. Let us check these two assumptions against the metric (\ref{2.7}).
We see that at $\nu=0$ it becomes
\begin{equation}
g_{\xi\xi}=a^2\cosh^2\xi,
\quad g_{\varphi\varphi}= a^2\sinh^2\xi,\quad g_{\varphi\xi}=g_{\xi\varphi}=0.
\label{2.77}
\end{equation}
Upon introducing a new variable $r=a\sinh\xi$ one gets
\begin{equation}
g_{rr}=g_{\xi\xi}(r)(\frac{\partial\xi}{\partial r})^2 = 1,
\quad g_{\varphi\varphi}= r^2,\quad g_{\varphi r}=g_{r\varphi}=0,
\label{2.777}
\end{equation}
which is nothing but the metric of a plane. This result is quite reasonable,  since putting $\nu=0$
should result in the trivial solution to the von Karman equations, $f=\chi=0$.
In the second case of interest, we have $\nu\neq 0$ and $a\gg 1$.
With the help of the above mentioned change of variables,
we get
\begin{equation}
g_{rr}= 1+(\frac{c}{a})^2\frac{r^2}{a^2+r^2},\quad
g_{\varphi\varphi}= \alpha^2r^2,\quad g_{\varphi r}=g_{r\varphi}=0.
\label{2.777}\end{equation}
We see that for large enough $a$, which in view of our assumption implies large $K_0$,
the elasticity properties come into play through the single
dimensionless parameter $\eta=c^2/a^2\ll 1$. In particular, in the inextensional limit $a\to \infty$, Eq.(\ref{2.777})
reduces to the metric of a cone.

We now argue that the desired form of $\eta$ follows directly from the von Karman equations.
To this end we suppose that there exist
two dimensionless functions $\tilde\chi$ and $\tilde f=f_0 +\tilde f$ that
fulfil Eqs. (\ref{3.16}) and at the same time account for the embedding (\ref{hyperb})
as well as metric (\ref{2.7}).
We have explicitly singled out the $f_0$ dependence to eliminate
the $\delta$-function source.
We are interested in the second "dynamical" equation in Eq. (\ref{3.16}). It takes the form
\begin{eqnarray}
\epsilon\Delta^2\tilde{\chi} &=& -K(f_0,\tilde f), \quad r\ge 1.
\label{3.24}\end{eqnarray}
To proceed we simply replace the Gaussian curvature in this equation with that of the
hyperboloid
\begin{equation}
K = \frac{c^2}{(a^2 +r^2(1+\eta))^2}.
\label{curvature}
\end{equation}
We recall that coordinate $r$ as well as parameters $a$ and $b$ are now dimensionless.
Equating then the both sides of (\ref{3.24}) at the core boundary $r=1$ yields
\begin{equation}
\eta^2=k c^2\epsilon +{\cal O}(\epsilon^2).
\label{3.25}\end{equation}
In getting this we have assumed that $a\gg 1$.
Here $k=\Delta^2\tilde{\chi}\mid_{r=1}$.
To restore an explicit $r_0$-dependence one needs to make the substitution $a\to a/r_0,\,
c\to c/r_0$. This yields
\begin{equation}
\eta^2=k(c/r_0)^2\epsilon.
\label{3.26}\end{equation}
It seems reasonable to assume that $c\sim \sqrt{\nu}r_0$,
which finally gives
\begin{equation}
\eta\sim \sqrt{\nu\epsilon}. \label{3.27}\end{equation}
Eq.~(\ref{3.27}) qualitatively relates the characteristics of
elastic media to the parameters of the embedding. It must be
stressed that the smooth hyperboloid parametrized by
Eq.(\ref{hyperb}) at $\nu=0$ is not a surface that replaces a rigid
cone for finite elasticity. Only when that hyperboloid of the near-zero
curvature is pierced by the gauge flux, the smoothed replacement of
the cone emerges. By construction $a\gg r_0$ and $c \sim r_0$, which
is consistent with $\eta\ll 1.$ This takes care of a small
stretching in the system.

Turning back to fermions, we see that the metric (\ref{2.7}) generates the spin connection term
\begin{equation}
\omega^{12}_{\varphi}=
-\omega^{21}_{\varphi}=
\left[1-\frac{a\alpha\,\cosh\xi}{\sqrt{g_{\xi\xi}}}\right]=\omega_{\alpha}(\xi).
\label{2.8}\end{equation}
Since $\omega_{\alpha}(\xi)\to 1-\alpha$ as $\xi\to 0,$ it in contrast with (\ref{2.5}) contains
a topologically nontrivial part that gives rise to a fixed
flux,
$$\lim_{\epsilon\to 0}\oint_{C_{\epsilon}}\omega^{12}_\varphi d\varphi=2\pi\nu.$$
We thus get the smoothed apex, the cone-like asymptotic at
large distances and the unremovable conical singularity at the
disclination line. It is known that in case a spin connection
contains an $SO(2)$ piece with nontrivial flux, that field cannot be
eliminated under any smooth deformation of the underlying metric
(see, e.g., \cite{witten}). Within our approach this simply means
that a nontrivial contribution to the spin connection which comes
from the topological gauge field survives any smooth elastic deformations of
the media.

\section{Flexible graphene}

In this section, we apply the developed approach to
describe the electronic properties of graphene with
pentagonal defects. Indeed, elastic characteristics of graphene are
well-fitted to our theory. The estimated bending rigidity of
graphene lies in the range of 1-2 eV and anyway does not exceed the
value of 2.5 eV (see, e.g., \cite{kudin,zakharchenko,wang}). At the
same time, the lower range value of $K_0r_0^2$ is approximately
given by 20 eV at $r_0=a_0$ with $a_0$ being the interatomic spacing
in graphene lattice~\cite{kudin}. Therefore, the parameter
$\epsilon$ is estimated as $\epsilon\leq 0.1$ thus
justifying an applicability of
the elasticity-induced perturbation scheme to graphene.

\subsection{Uniform magnetic field: Landau states}

The Dirac equation on a surface
$\Sigma$ in the presence of the gauge field $a_{b}$ and the
external magnetic field with the vector potential $A_{b}$ is
written as
\begin{equation}
i\gamma^{\alpha}e_{\alpha}^{\ b}[\nabla_{b} - ia_{b}-
iA_{b}]\psi = E\psi, \label{eq:4.1}
\end{equation}
where $\nabla_{b}=\partial_{b}+\Omega_{b}$. The effective abelian gauge
field $a_{b}$ is responsible for valley mixing since $K$ and $K'$ points
become inequivalent in the presence of the pentagonal defect
(see, e.g.,~\cite{review} for detail). The
energy in (\ref{eq:4.1}) is measured from the Fermi level.

On a surface of the hyperboloid the Dirac operator reads
\begin{equation}
\hat{D}=\left(\begin{array}{cc}0&e^{-i\varphi}\big(-\frac{\partial_{\xi}}{\sqrt{g_{\xi\xi}}}+
\frac{1}{a\alpha\sinh\xi}(i\partial_{\varphi}+\frac{1}{2}\omega_{\alpha}(\xi)+\Omega_{\varphi})\big)\\
e^{i\varphi}\big(\frac{\partial_{\xi}}{\sqrt{g_{\xi\xi}}}+\frac{1}{a\alpha\sinh\xi}(i\partial_{\varphi}-
\frac{1}{2}\omega_{\alpha}(\xi)+\Omega_{\varphi})\big)&0\end{array}\right),
\label{eq:4.3}\end{equation} where
\begin{equation}
\Omega_{\varphi}=a_{\varphi}+A_{\varphi}\nonumber.
\end{equation}
It can be verified that $\hat{D}=\hat{D}^{\dag}$. The substitutions
\begin{equation}
\left(%
\begin{array}{c}
  \psi_{A} \\
  \psi_B \\
\end{array}%
\right) =\frac{1}{\sqrt{2\pi}}\left(%
\begin{array}{c}
  u(\xi)e^{ij\varphi} \\
  v(\xi) e^{i(j+1)\varphi}\\
\end{array}%
\right),\quad j=0,\pm1,...,\label{eq:4.4}
\end{equation}
and
\begin{equation}
\tilde{\psi}=\psi\sqrt{\sinh\xi},\label{eq:4.5}
\end{equation}
reduce the Dirac equation $\hat{D}\psi=E\psi$ to the form
\begin{equation}
\partial_{\xi}\tilde{u}-\frac{(j+1/2-a_{\varphi}+A_{\varphi})}{\alpha}\sqrt{\coth^{2}\xi+\eta}\tilde{u}
=\tilde{E}\tilde{v},\label{eq:4.6}
\end{equation}
\begin{equation}
-\partial_{\xi}\tilde{v}-\frac{(j+1/2-a_{\varphi}+A_{\varphi})}{\alpha}\sqrt{\coth^{2}\xi+\eta}\tilde{v}
=\tilde{E}\tilde{u}, \label{eq:4.7}
\end{equation}
where
$\tilde{E}=\sqrt{g_{\xi\xi}}E$.
Explicitly, $a_{\varphi}=\pm(N/4+M/3)$ for even number of
pentagons and $a_{\varphi}=\pm N/4$ for odd number of pentagons.
Here $N$ is the number of pentagonal defects and $M$ takes the
values $M=-1,0,1$ depending upon the arrangement of pentagons
(see~\cite{Crespi} for detail).
The uniform external magnetic field $B$ is chosen to be pointed in
the $z$-direction so that $ \vec{A}=B\left(y,-x,0\right)/2$. In
$(\xi$,  $\varphi)$ coordinates the components of $A_b$ are
written as
\begin{equation}
A_{\varphi}=-\Phi\sinh^{2}\xi;\quad
A_{\xi}=0,
\label{eq:7}
\end{equation}
where $\Phi=ba^{2}/2$ and $b=eB/\hbar c$. Let us start with the
analysis of the electron state at the Fermi level (so-called
zero-energy mode). In this case, we put E=0 in (\ref{eq:4.6}) and
(\ref{eq:4.7}). The exact solution is found to be
\begin{eqnarray}
\tilde{u_{0}}(\xi)=C\left(\Delta+k\cosh\xi\right)^{k\tilde j+\frac{\eta\tilde\Phi}{2k}}
\left(\frac{\Delta+\cosh\xi}{\sinh\xi}\right)^{-\tilde j}
\exp\left({-\frac{\tilde\Phi\Delta\cosh\xi}{2}}\right),
\label{eq:4.8}
\end{eqnarray}

\begin{eqnarray}
\tilde{v_{0}}(\xi)=C'\left(\Delta+k\cosh\xi\right)^{-k\tilde j-\frac{\eta\Phi}{2k}}
\left(\frac{\Delta+\cosh\xi}{\sinh\xi}\right)^{\tilde{j}}
\exp\left(\frac{\tilde\Phi\Delta\cosh\xi}{2}\right),
\label{eq:4.9}
\end{eqnarray}
where
$$
k=\sqrt{1+\eta};\quad
\Delta=\Delta(\xi)=\sqrt{1+k^{2}\sinh^{2}\xi},\quad \tilde j=(j+1/2-a_{\varphi})/\alpha,\quad \tilde\Phi=\Phi/\alpha,
$$
and $C$ and $C'$ are the normalization factors.
Evidently, the only component $u_0$ becomes normalizable.
In the inextentional limit ($\eta\rightarrow 0$)
one obtains
\begin{equation}
u_0(r)\propto
r^{\tilde j-1/2}\exp\left({-\frac{b r^2}{4\alpha}}\right),
\label{eq:4.10}
\end{equation}
where $r=a\sinh\xi$. This result agrees with that of~\cite{Crespi}.

Let us consider the Landau states. To this end, we develop the
perturbation scheme using $\eta\sim\nu\epsilon$ as the perturbation parameter.
Indeed, according to (\ref{2.777}) the hyperboloid
can at small $\eta$ be considered as a local perturbation of the cone metric. Notice
that a similar procedure was used in the description of electronic
states in spheroidal fullerenes~\cite{Pudlak,Pudlak1} where the
spheroid was considered as a slightly elliptically deformed sphere.
At the same time, the Landau states on the graphene
cone were already studied in detail in~\cite{Crespi}. Therefore, one
can use the unperturbed solutions found there.

The Dirac operator is written as
\begin{equation}
\hat{\cal{D}}=\hat{\cal{D}}_{0}+\eta\hat{\cal{D}}_{1},
\label{eq:4.11}\end{equation}
where
$$\hat{\cal{D}}_{0}=i
\gamma_{2}\frac{1}{a\cosh\xi}\left
(\partial_{\xi}+\frac{\cosh\xi}{2\sinh\xi}\right
)-\frac{\gamma_{1}}{a\sinh\xi}\left
(\tilde j+\tilde\Phi\sinh^{2}\xi\right )
$$
is the Dirac operator on the true cone. It is important to note here
that generally the operator $\hat{\cal{D}}_{1}$ is not Hermitian on
a cone and must be extended to a Hermitian one (see, e.g., the discussion in~\cite{Pudlak,Pudlak1}).
The result is
$$
\hat{\cal{D}}_{1}=
-\frac{\gamma_{1}\sinh\xi}{2a\cosh^{2}\xi}\left(\tilde j+\tilde\Phi\sinh^{2}\xi\right).
$$
It is convenient to square the Dirac operator,
\begin{equation}
\hat{\cal{D}}^{2}=\hat{\cal{D}}_{0}^{2}+\eta\hat{\Gamma},
\label{eq:4.12}\end{equation} where
$\hat{\Gamma}=\hat{\cal{D}}_{0}\hat{\cal{D}}_{1}+\hat{\cal{D}}_{1}\hat{\cal{D}}_{0}$ and
the quadratic in $\eta$ term is omitted.
Explicitly,
\begin{equation}
\hat{\Gamma}= \frac{1}{a^{2}\cosh^2\xi}\left[\left(\tilde j+
\tilde\Phi\sinh^{2}\xi\right)^2
+\frac{\sigma_{3}}{2}\left(\tilde j\left(\frac{2}{\cosh^{2}\xi}-1\right)+
\tilde\Phi\sinh^{2}\xi\left(\frac{2}{\cosh^{2}\xi}+1\right)\right)\right].
\label{eq:4.13}\end{equation} For $\eta=0$ both the unperturbed wave
functions $\psi_{jn}$ and the Landau energy levels were obtained
in~\cite{Crespi}. Two families of solutions were found. We restrict
our consideration here to the first family where $\tilde j\geq0$. In this
case, the energy of the so-called bulk levels reads~\cite{Crespi}
$E^{0}_n=\pm\sqrt{2n}$, $n=0,1,2...,$ where the energy is measured in
units of $\hbar v_F/l_B$ with the magnetic length $l_B=\sqrt{\hbar
c/eB}$.

Let us calculate the matrix element of the perturbation $\Gamma=\langle\psi_{jn}|\hat{\Gamma}|\psi_{jn}\rangle$.
Our analysis shows that the perturbation does not influence the zero energy level
in the first order in $\eta$. Since the resulting expressions are rather involved,
we calculate the first energy level numerically. The first Landau level is found to be
slightly shifted due to the elastic contribution,
\begin{equation}
E^{\eta}_{n=1}\simeq \pm\sqrt{2}\pm\eta\frac{\Gamma}{2\sqrt{2}}\simeq\pm\sqrt{2}\pm 0.3\eta,\quad \eta>0,
\label{eq:4.14}
\end{equation}
where the $\pm$ sign corresponds to the conduction and valence band, respectively.
Note that a similar shift of the first Landau level follows
for the second ($\tilde j\leq 0$) family of the solutions referred to
as the apical states~\cite{Crespi}.

In graphene there are many allowed transitions due to the presence of two electron bands, the conduction
and the valence band, and the transitions have the energies
\begin{equation}
\Delta_{n+1,n}^{\xi}=\frac{\hbar v_F}{l_B}[\sqrt{2(n+1)}-\xi\sqrt{2n}],
\label{trans}\end{equation}
where $\xi=\pm$ denote the intraband and interband transitions, respectively~\cite{goerbig}.
As is seen from Eq.(\ref{eq:4.14}) the transition between the ground energy level and the first one
in the presence of elasticity is modified to become
\begin{equation}
\Delta_{1,0}=\frac{\hbar v_F}{l_B}[\sqrt{2}+0.3\eta].
\label{trans1}\end{equation}
The experimentally observed cyclotron
resonancelike and electron-positron-like transitions are in a good
agreement with the theoretical expectations of a single-particle
model of Dirac fermions in graphene~\cite{sadowski}. They produce
a very accurate value for $v_F$, the velocity of electrons in
graphene. In our case, the Fermi velocity becomes
slightly renormalized due to the elasticity effects induced by
the phenomenological parameter
$\eta: \Delta v_{F}/v_F=0.3\eta/\sqrt{2}$.
Therefore, it would be interesting to perform
analogous experiments with graphite cones instead of graphene
to determine the modification of $v_F$.

\subsection{Zero magnetic field: Electronic density of states}

Let us consider now the case of zero magnetic field, $B=0$.
This markedly changes the situation with the zero-mode states.
Indeed, in the absence of the cut-off exponent in (\ref{eq:4.8})
and (\ref{eq:4.9}) the normalization conditions are found to be
$-1/2<\tilde j<-1/2k$ for $u_0(\xi)$
and $1/2k<\tilde j<1/2$ for $v_0(\xi)$.
As a result, at small $\eta$ which is of interest here, there are no
normalized solutions. This means that smoothing has no marked effect
on the existence of zero modes. As stated above,
one of the modes (either $\tilde u(\xi)$ or $\tilde v(\xi)$)
becomes in the presence of
the uniform magnetic field directed along the $z$-axis
nonmailable and there exists a true zero mode.
Therefore, one can expect a "switching-like" effect
driven by the magnetic field.

An interesting question is how a hyperboloid geometry influences
the density of states near the Fermi energy in the vicinity of the
pentagonal defects. We will be interested in the DOS in a small
ring $0<r\leq\delta$ around the defects. First, we need to
find corrections to the wave functions for the hyperboloid
geometry. Following the perturbation scheme with $\eta$ being a
small parameter, one can write
$\psi_n=\psi^{0}_n+\psi^{\eta}_n+...$ with $\psi^{0}_n$ being the
solutions for $\eta=0$ which are the ordinary Bessel functions
(see~\cite{lammert}). Here
$\psi^{\eta}_n\propto\sum_{m}\Gamma_{mn}\psi_{m}^{0}/\Delta k$ is
a perturbative part with $\Delta k=\pi/R$. The matrix element of
the perturbation reads $\Gamma_{mn}=
\langle\psi_{jm}^{0}|\hat{\Gamma}|\psi_{jn}^{0}\rangle$, where the
perturbation term takes the form
\begin{equation}
\hat{\Gamma}= \frac{\tilde j}{a^{2}\cosh^{2}\xi}\left[\tilde j
+\frac{\sigma_{3}}{2}\left(\frac{2}{\cosh^{2}\xi}-1\right)\right].
\label{eq:4.15}
\end{equation}
Finally, the total DOS in the $\delta$ disk is found to be
\begin{equation}
D(E,\delta,n)\propto DOS^{0}+\eta DOS^{\eta}\propto
DOS^{0}(1+\eta|\frac{E}{a}|\delta^{2}),\quad \eta>0, \label{eq:4.16}
\end{equation}
where
$$
DOS^{0}\propto|E|^{2n+1}\delta^{2(n+1)},
$$
coincides with the total DOS near the defects for graphitic cones
found in~\cite{lammert}, and $n=\tilde j\pm1/2$ is the index of the
Bessel function. The term
\begin{equation}
DOS^{\eta}\propto|E|^{2(n+1)}\delta^{2(n+2)} \label{eq:4.17},
\end{equation}
comes from the elastic perturbation. Thus, we obtain an increase of DOS
in the vicinity of the defect due to the finite elasticity. This behavior
can be explained by the elasticity-induced contribution to the
underlying metric which markedly deviates from that of a true cone near the apex.
The modification of the DOS is more pronounced for more flexible membranes.
Notice that the bending rigidity of square graphene is found to grow with
a size~\cite{wang} while the Young's modulus depends on the temperature~\cite{zakharchenko}.
These findings give a possibility of experimental studies of the influence of elasticity
on the electronic characteristics of disclinated graphene. As a promising material, one can
consider the graphene monolayers obtained via chemical reduction of graphene oxide which
have excellent mechanical properties including high bending flexibility and tensile strength~\cite{kern}.

\section{Conclusion}

In conclusion, we have presented an analytical approach to
describe Dirac fermions on a flexible disclinated surface
beyond the inextensional limit.
The elastic membrane
is considered as an embedding of a $2D$ surface into $R^3$. The disclination
is incorporated through a topologically nontrivial $SO(2)$  gauge field
that generates a metric with a conical
singularity. A smoothing of the conical singularity
is accounted for by regarding the upper half of a disclinated
two-sheet hyperboloid as an elasticity-induced embedding.
Parameters of that embedding are chosen to match
the solution to the von Karman equations.
Away from but close to the inextensional limit,
the Young's modulus $K_0$ and bending rigidity $\kappa$ enter the theory
through a dimensionless parameter, $\epsilon=\kappa/(K_0r_0^2)\ll 1$,
where $r_0$
sets the relevant short-range scale: a radius of the defect core.

We argue that the homogeneous part of the solution to
the von Karman equations for disclinated membrane is
of the utmost importance in stabilizing the theory.
This finding allows us to avoid
the evident problem with the core radius mentioned in~\cite{nelson}.
Surprisingly, the discovered homogeneous part of the solution
has been missed so far.
We calculate the stretching energy of the membrane to observe
the logarithmic behavior with the membrane radius $R$ similar to
that of the bending energy. For rigid enough membranes
we work out the self-consisting perturbation scheme with $\epsilon$
being the small parameter.

We apply a new approach to study the structure of the low energy electronic
states of flexible graphene with a topological defect. We find a true zero-mode
state in the presence of an external uniform magnetic field.
The finite elasticity results in a smearing of the cone apex thus
modifying the cone metric. The elasticity affects enter the observable quantities
through the elasticity-induced phenomenological parameter $\eta \sim \sqrt{\nu\epsilon}$.
Qualitatively, the first Landau level in the conduction band is found to be shifted upwards,
whereas the corresponding one in the valence band is shifted downwards.
The total DOS near the tip is shown to increase due to elasticity.

From the experimental point of view,
it would be interesting to carry out experiments with elastic graphite cones
to determine the modification of $v_F$ due to elastic deformations caused by a
disclination defect.
The elasticity effects are more significant
for more flexible materials. Therefore graphene-based materials with excellent
mechanical properties like those recently obtained via chemical reduction of graphene oxide
in~\cite{kern} would be of interest in these studies.

\vskip 0.2cm ACKNOWLEDGEMENTS ---E.A.K. and V.A.O. acknowledge
financial support by the Russian Foundation for Basic Research
under grant No. 08-02-01027. R.P. acknowledges financial support by
the Slovak Academy of Sciences in the framework of CEX NANOFLUID,
and by the Science and Technology Assistance Agency under Contract
No. APVV 0509-07 and by VEGA Grant No. 2/0069/10.

\end{document}